\documentclass[onecolumn,notitlepage,nofootinbib]{revtex4-1}

\usepackage{graphicx}			
\usepackage{amssymb,amsfonts,amsmath}
\usepackage[font=small]{caption} 
\usepackage[usenames,dvipsnames]{xcolor}
\usepackage[colorlinks=false,linkcolor=blue,urlcolor=blue,citecolor=blue]{hyperref}
\usepackage{fullpage}

\newcommand{\pc}{p_\mathrm{c}}
\newcommand{\fc}{f_\mathrm{c}}
\newcommand{\xn}{\lceil n\fc\rceil}
\newcommand{\coloronline}{}

\def\lett#1{(\textbf{#1})}

\newcommand{\etal}{\emph{et al}.\@}

\providecommand{\shortcite}[1]{\cite{#1}}

\begin{document}

\title{Robustness and modular structure in networks}
\date{July 13, 2015}

\author{James P.~Bagrow}
\affiliation{Mathematics \& Statistics, University of Vermont, Burlington, VT, USA}
\affiliation{Center for Complex Network Research, Northeastern University, Boston, MA, USA}
\author{Sune Lehmann}
\affiliation{DTU Informatics, Technical University of Denmark, Kgs Lyngby, Denmark}
\affiliation{College of Computer and Information Science, Northeastern University, Boston, MA, USA}
\author{Yong-Yeol Ahn}
\affiliation{School of Informatics \& Computing, Indiana University, Bloomington IL, USA}
\affiliation{Center for Complex Network Research, Northeastern University, Boston, MA, USA}

\begin{abstract}
Complex networks have recently attracted much interest due to their prevalence
in nature and our daily lives~\cite{vespignani2009predicting,newman2010}.  A
critical property of a network is its resilience to random breakdown and
failure~\cite{albert2000,PhysRevLett.85.4626,
PhysRevLett.85.5468,PhysRevLett.86.3682}, typically studied as a percolation
problem~\cite{Stauffer_Aharony_1992,Achlioptas13032009,PhysRevLett.106.115701}
or by modeling cascading
failures~\cite{PhysRevLett.93.098701,buldyrev-havlin-catastrophic-cascade-failures-nature-2010,Brummitt20032012}.                       
Many complex systems, from power grids and the Internet to the brain and
society~\cite{colizza2007reaction,vespignani2011modelling,balcan2011phase}, can
be modeled using modular networks comprised of small, densely connected groups
of nodes~\cite{Girvan11062002,citeulike:686555}.  These modules often overlap,
with network elements belonging to multiple
modules~\cite{palla05,ahnbagrowlehman10}.  Yet existing work on robustness has
not considered the role of overlapping, modular structure.
Here we study the robustness of these systems to the failure of elements.  We
show analytically and empirically that it is possible for the modules
themselves to become uncoupled or non-overlapping well before the network
disintegrates.  If overlapping modular organization plays a role in overall functionality,
networks may be far more vulnerable than predicted by conventional percolation
theory.  
\end{abstract}

\maketitle

\section{Introduction}

Consider a system of interacting \emph{elements} representing computers, power
generators, neurons, office workers, etc.  Typically these elements fulfill individual roles in
the network such as regulating power or propagating neuronal signals. Yet in
many systems, global functionality may require elements to also perform
collective tasks of sufficient complexity that they cannot be completed by
single elements. The elements must instead work closely in teams, forming
densely interconnected \emph{modules}.  
These tasks may be parallelised computations, protein biosynthesis, or
higher-order neurological functions such as visual processing or speech
production.

Among various ways that modules can communicate or jointly function with each other, a prominent one is sharing common elements across modules.
Such overlapping elements may result from, e.g.,
pleiotropy in the genome~\cite{StearnsPleitropy}  or ``structural
folds'' in social systems~\cite{vedrasStructuralFolds2010}. 
In biological networks, functional modules often coordinate by sharing common elements. For instance, the recent yeast protein complex catalogue, with high-quality hand-curated protein complex data, showed that a significant fraction of proteins belong to multiple protein complexes~\cite{Pu01022009}.
In functional brain networks, overlap may indicate regions that integrate, e.g.,
visual and auditory sensory cues~\cite{kaiser2011tutorial}, while in structural brain networks it was shown that ``confluence zones'' that integrate information from other regions tend to participate in multiple modules~\cite{de2014edge}.
In large human organizations, liaison jobs---where workers coordinate cross-team activities by spending significant time
in multiple teams---are common~\cite{liasonRoles1974}. 
For instance, many companies implement hierarchical management systems, where a manager oversees several teams. In such cases, coordination between the teams are arranged mainly by the manager who works with all the teams, playing the role of the overlapping element. Moreover, collaboration and management across multiple locations frequently takes the form of swapping or dispatching personnel across places.
Another example is the systems analyst~\cite{systemsAnalyst1982}. Systems analysts help organizations
improve their information technology offerings by liaising between end users or external vendors outside the organization and the programming teams of the organization itself.  Similarly, the militaries of  many nations have command structures dedicated
to liaising between different military branches during joint operations, including with the militaries of other countries~\cite{vego2009joint}.

Inspired by these examples,
we ask how these networks respond when a random fraction of
elements fail: can the modules become uncoupled (i.e., non-overlapping) before the network loses global
connectivity?  Random failures provide a general model of, e.g., a traumatic brain
injury or degenerative disease.  If enough elements fail, overlap may be lost and some or all modules may no
longer be able to complete their tasks (higher brain functions are lost) even though the network may
remain connected (simpler autonomic responses persist).  Likewise, an individual
module may fail if too many of its member elements cease to function. These effects in combination
may lead to a loss of modular overlap in the system, which by our simplified assumptions causes impairment to the entire system. 
See Fig.~\ref{fig:cartoon} for an example illustrating how the loss of a random element may cause a module
to fail.

The rest of this paper is organized as follows. In Sec.~\ref{sec:models} we study the robustness of an analytically tractable model
of modular networks (Secs.~\ref{subsec:elementnetwork}, \ref{subsec:modulenetwork}), as well as additional models of modules (Sec.~\ref{subsec:additionalModelsModules}) and models of how those modules may
respond to random failures and targeted attacks (Sec.~\ref{subsec:additionalModelsOfModuleFailures}). 
Section~\ref{sec:empirical} supports these results with studies of four
empirical network datasets covering very different research domains. Finally, in Sec.~\ref{sec:discuss} we discuss the context of this work and how it may apply to the practical issue of missing data during the detection of 
overlapping and non-overlapping communities in real-world network datasets.

\section{Modeling modular networks}
\label{sec:models}

\begin{figure}
    \centering
    \includegraphics[trim=0 00 0 0,clip=true,width=0.75\textwidth]{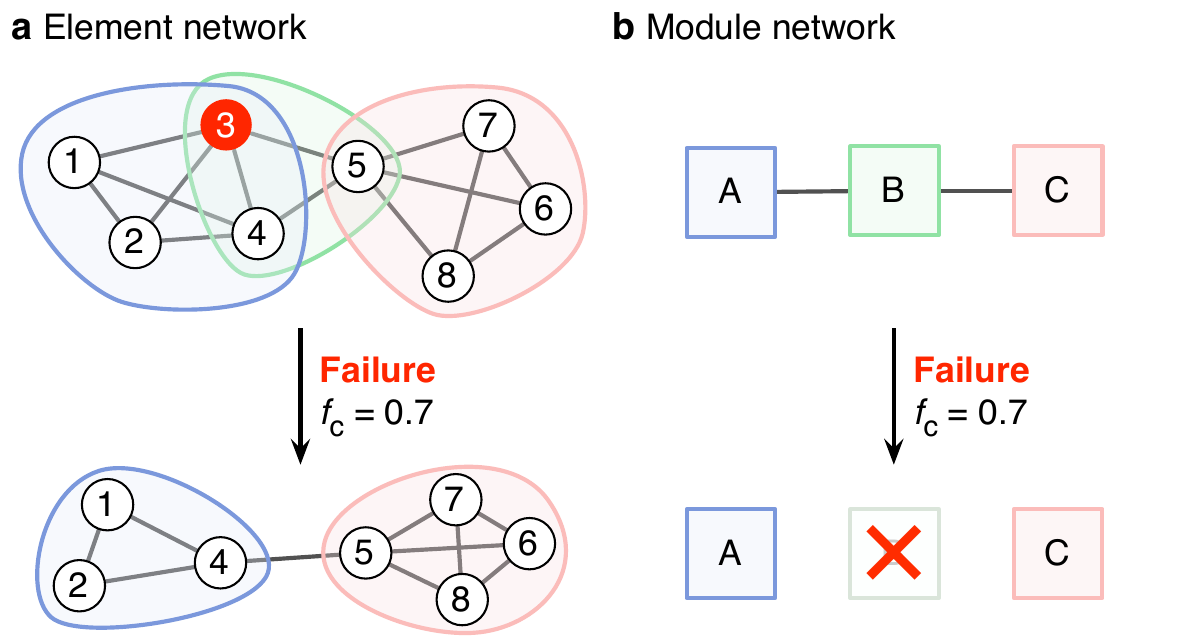}
    \caption{Modeling failures in modular networks.
        We analyse two networks, one representing the linkages between network
        elements \lett{a} and a second detailing the overlapping connectivity
        between the modules themselves \lett{b}.
        In this example, the failure of element 3 leads to the loss of module B,
        since B no longer has sufficient members to complete its collective
        task. This causes the module network to become disconnected (bottom)
        even though the element network remains connected.
    \label{fig:cartoon}}
\end{figure}

Networks with overlapping modular structure can be well modeled with a bipartite graph, also known as an affiliation network~\cite{wasserman1994social}. This network consists of two types of nodes representing the elements and the modules and undirected links representing which elements belong to which modules. 
Links in the bipartite graph only connect element nodes to module nodes.
The network is
characterized by two degree distributions, $r_m$ and $s_n$, governing the
fraction of elements that belong to $m$ modules and the fraction of modules that
contain $n$ elements,
respectively~\cite{newman2002random,PhysRevE.64.026118,newman,newmanpark}. 
Links are placed randomly between element and module nodes respecting these degree distributions~\cite{newmanpark}.
The
average number of modules per element is $\sum_m m r_m \equiv \mu$ and the
average number of elements per module is $\sum_n n s_n \equiv \nu$.
Using this as a starting point for our model, we derive two networks from the
bipartite graph by projecting onto either the elements or the modules: One is
the network between elements, studied by Newman \shortcite{newman} and Newman \& Park 
\shortcite{newmanpark}, while the other is a network where each node
represents a module and two modules are linked if they share at least one
element.  The Largest Connected Component (LCC) (also known as the giant component~\cite{Stauffer_Aharony_1992}) in the element network
disappears when, due to missing elements, the network loses global connectivity; in the module network it
vanishes if the modules become uncoupled (non-overlapping). 
Before projection elements fail independently with probability $1-p$ and are removed from the
network.  Meanwhile, a module is unable to complete its collective task if fewer
than a critical fraction $\fc$ of its original elements remain.  These failed
modules are removed from the module network but any surviving member elements
are not removed from the element network (Fig.~\ref{fig:cartoon}). 
This model of element and module failure serves as our starting point, but we will also explore additional models.

Before we analyze this model, it is important to note that it makes two
assumptions about the modular nature of the system: that all interactions within
each module exist and are equal, and that there are no differences between
individual elements that share a module---i.e., there are no ``captains'' or
``team leaders''. One may expect that these homogeneous (or ``mean-field'') assumptions
may limit the applicability of results derived from this model. 
However, we argue that, when large numbers of modules are involved, such
microscopic, per-module details are less important to the macroscopic robustness
of the large-scale system than the overall organization of the network's modules. That is, an averaged or mean-field model for module
failure captures the most essential elements of many systems' robustness.
Furthermore, we will present a number of findings that relax these mean-field
ingredients.

We wish to determine $S(p)$, the fraction of remaining nodes within the LCC as a function of $p$, for both the element and module networks.  We
use four generating functions~\cite{newman,newmanpark}:
\begin{equation}
    \begin{aligned}
        f_0(z) &= \sum_{m=0}^{\infty} r_m z^m, & f_1(z) &= \frac{1}{\mu}\sum_{m=0}^\infty m r_m z^{m-1},\\
        g_0(z) &= \sum_{n=0}^{\infty} s_n z^n, & g_1(z) &= \frac{1}{\nu}\sum_{n=0}^\infty n s_n z^{n-1}.
    \end{aligned}
\end{equation}
These functions generate the probabilities for ($f_0$) a randomly chosen element
to belong to $m$ modules, ($f_1$) a random element within a randomly chosen
module to belong to $m$ other modules, ($g_0$) a random module to contain $n$
elements, and ($g_1$) a random module of a randomly chosen element to contain
$n$ other elements.

To analyse this model we now separately study the two projections (the element and module networks) of the original bipartite graph.

\subsection[]{Element network\protect\footnote{This short calculation was presented in Newman \shortcite{newman} and Newman \& Park \shortcite{newmanpark}. We repeat it here for completeness and to introduce notation used for subsequent calculations.}}
\label{subsec:elementnetwork}

Consider a randomly chosen element A that belongs to a group of size $n$.  Let
$P(k|n)$ be the probability that A still belongs to a connected cluster of $k$
nodes (including itself) in this group after failures occur:
\begin{equation}
P(k|n) = \binom{n-1}{k-1} p^{k-1}(1-p)^{n-k}.
\end{equation}
The generating function for the number of other elements connected to A within
this group is
\begin{equation}
    h_n(z) = \sum_{k=1}^n P(k|n) z^{k-1} = \left(z p + 1-p\right)^{n-1}.
\end{equation}
Averaging over module size:
\begin{equation}
    h(z) = \frac{1}{\nu}\sum_{n=0}^{\infty} n s_n h_n(z) = g_1(z p + 1-p).
\end{equation}
The total number of elements that A is connected to, from all modules it belongs
to, is then generated by
\begin{equation}
    G_0(z) = f_0(h(z)).
    \label{eqn:G0}
\end{equation}
Likewise, the total number of elements that a randomly chosen neighbor of A is
connected to is generated by
\begin{equation}
    G_1(z) = f_1(h(z)).
    \label{eqn:G1}
\end{equation}

Before determining $S$, we first identify the critical point $\pc$ where the
giant component emerges.  This happens when the expected number of elements two
steps away from a random element exceeds the number one step away, or
\begin{equation}
    \partial_z G_0(G_1(z))\big|_{z=1} - \partial_z G_0(z)\big|_{z=1} > 0.
\end{equation}
Substituting Eqs.~\eqref{eqn:G0} and \eqref{eqn:G1} gives
$f_0'(1)h'(1)[f_1'(1)h'(1)-1] > 0$ or $f_1'(1)h'(1) > 1$.  Finally, the
condition for a giant component to exist, since $h'(1) = p g_1'(1)$, is
\begin{equation}
 p f_1'(1) g_1'(1) > 1.
\end{equation}
For the uniform case, $r_m = \delta(m,\mu)$ and $s_n = \delta(n,\nu)$, this
gives $p (\mu-1) (\nu-1) > 1$.  If $\mu=3$ and $\nu=3$, then the transition
occurs at $\pc = 1/4$.

To find $S$, consider the probability $u$ for element A not to belong to the
giant component. A is not a member of the giant component only if all of A's
neighbors are also not members, so $u$ satisfies the self-consistency condition
$u = G_1(u)$.  The size of the giant component is then $S = 1 - G_0(u)$.

\subsection{Module network}\label{subsec:modulenetwork}

Consider a random module C and then a random member element A.   Let $Q(\ell|m)$
be the probability that C is connected to $\ell$ modules, including itself,
through element A, who was originally connected to $m$ modules including C: 
\begin{equation}
    Q(\ell|m) = \binom{m-1}{\ell-1} q_1^{\ell-1} \left(1-q_1\right)^{m-\ell},
\end{equation}
where 
\begin{equation}
    q_1 = \frac{1}{\nu}\sum_{n=0}^\infty n s_n \sum_{i=x}^n \binom{n-1}{i-1} p^{i-1} (1-p)^{n-i}.
\end{equation}
(Notice that $q_1 = 1$ when $x(n) \equiv \xn = 1$ for all $n$.)
The generating function $j_m$ for the number of modules that C is connected to, including itself,
through A is
\begin{equation}
    j_m(z) = \sum_{\ell=1}^m Q(\ell|m) z^{\ell-1} = \left(z q_1 + 1-q_1\right)^{m-1}.
\end{equation}
Once again, averaging $j_m$ over memberships gives
\begin{equation}
    j(z) = \frac{1}{\mu} \sum_{m=0}^\infty m r_m j_m(z) = f_1(z q_1 + 1-q_1).
\end{equation}
The total number of modules that C is connected to is \emph{not} generated by $g_0(j(z))$ but by
$\tilde{g}_0(j(z))$, where the $\tilde{g}_i$ are the generating functions for module size after
elements fail:
\begin{align}
    \tilde{g}_0(z)   &= \sum_{n=0}^{\infty} \tilde{s}_n z^n, 
    & \tilde{g}_1(z) &= \frac{\sum_{n=0}^\infty n \tilde{s}_n z^{n-1}}{\sum_{n=0}^{\infty} n \tilde{s}_n}.
\end{align}
The probability $\tilde{s}_k$ to have $k$ member elements remaining in a module after percolation is
given by 
\begin{equation}
    \tilde{s}_k = \frac{\sum_n \binom{n}{k} p^k (1-p)^{n-k} s_n}{\sum_n \sum_{k'=x}^{n}
        \binom{n}{k'} p^{k'} (1-p)^{n-k'} s_n}.
\end{equation}
The denominator is necessary for normalisation since we cannot observe modules with fewer than $\xn$
members.  Notice that  $\tilde{s}_n = s_n$ when $s_n = \delta(n,\nu)$ and $\xn=n=\nu$. 

Finally, the total number of modules connected to C through any member elements
is generated by $F_0(z) = \tilde{g}_0(j(z))$ and the total number of modules
connected to a random neighbor of C is generated by $F_1(z) =
\tilde{g}_1(j(z))$.  As before, the module network has a giant component when
$\partial_z F_0(F_1(z))|_{z=1} - \partial_z F_0(z)|_{z=1} > 0$ and $S = 1-
F_0(u) = 1 - \tilde{g}_0(j(u))$, where $u$ satisfies $u = F_1(u) =
\tilde{g}_1(j(u))$.

\begin{figure}
    \centering
    \includegraphics[width=0.75\textwidth]{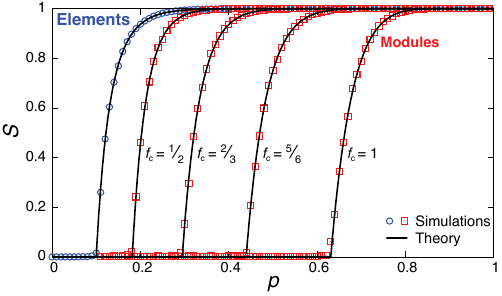}
    \caption{The size of the Largest Connected Component $S$ for the element and module networks.
        Theory and simulations confirm that the network undergoes a transition
        from coupled to non-overlapping modules well before it loses global
        connectivity.  Symbols represent element ($\odot$) and module
       ($\boxdot$) networks.  
       Here we used $r_m = \delta(m,\mu)$, $s_n =
        \delta(n,\nu)$, with $\mu=3$ and $\nu=6$. Simulations used $10^6$
        elements.
    \label{fig:results}}
\end{figure}

For the uniform case with $\mu = 3$, $\nu = 3$, and $\fc>2/3$, the critical
point for the module network is $\pc = 1/2$, a considerably higher threshold
than for the element network ($\pc = 1/4$).  In Fig.~\ref{fig:results} we show
$S$ for $\mu=3$ and $\nu=6$.  The \emph{robustness gap}, the difference between the critical points for the element and
module networks, grows as the module failure cutoff increases, covering a
significant range of $p$ for the larger values of $\fc$.

\begin{figure}
    \centering
    {\includegraphics[width=0.75\textwidth,trim=0 14 0 17,clip=true]{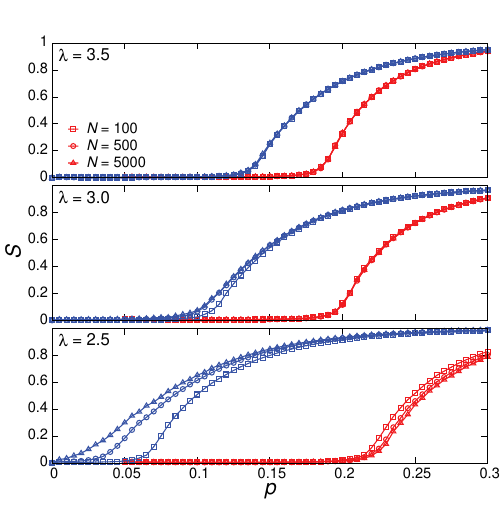}}
    \caption{Robustness of scale-free modular networks. %
        Here $r_m = \delta(m,3)$, $s_n \sim n^{-\lambda}$, $\fc = 1/2$, and $N
        \equiv \max \{n \mid s_n > 0\}$.  Increasing $N$ and decreasing
        $\lambda$, measures known to improve the robustness of scale-free
        networks~\protect\cite{PhysRevLett.85.4626}, actually magnifies the difference
        between the critical points.  Surprisingly, this also increases the
        fragility of the module network, indicating that optimizing against
        structural failure may worsen the network's functional resilience.
        Simulations used $10^5$ elements.\label{fig:scalefree}}
\end{figure}

Of particular interest are scale-free networks~\cite{BarabasiRekaEmergence1999,newman2010}.
Here we take $r_m = \delta(m,\mu)$ as before, but now $s_n \sim n^{-\lambda}$, with $\lambda \geq
2$. 
(The degree distribution after projection remains scale-free, with the same
exponent, although the maximum degree may increase.)
It is known that scale-free networks are robust to
random failures when $2 < \lambda < 3$ (meaning that $\pc \to 0$).  However, this result also
requires that the maximum value $K$ of the degree distribution be large ($K \gg
1$)~\cite{PhysRevLett.85.4626}.
Indeed, as we lower $\lambda$, we discover that, while we increase
the robustness of the elements, we actually \emph{decrease} the robustness of the modules
(Fig.~\ref{fig:scalefree}).
Interestingly, increasing the maximum module size cutoff $N = \max\{n \mid s_n > 0\}$ 
improves element robustness, but not overall functional resilience.

\subsection{Additional models of modules}\label{subsec:additionalModelsModules}

Our analytic calculation (Secs.~\ref{subsec:elementnetwork} and \ref{subsec:modulenetwork}) uses a basic, mean-field model of modular structure.
Specifically, we follow Newman and Park~\cite{newmanpark} and represent the element-element
network as the \emph{projection} of a bipartite graph between elements and
modules.  This assumes that each module is fully dense, i.e.\ that all
interactions within the module are present and of equal strength (hence the
mean-field nature). Due to these assumptions, this network model is tractable,
a great strength.
Yet while we expect modules to be unusually dense, it is unlikely that they will
universally be \emph{completely} dense.   However, when considering expected
behavior over many modules, which is the primary factor of the model's global network
robustness, we argue that such higher-order effects and potential microscopic
details are subordinate to gross modular features when studying global network connectivity.

Despite this, understanding how more detailed modular representations affect
network robustness is important. As a first step we relax our assumption of
mean-field intra-modular coupling. Before elements fail (site percolation) we
delete each link in the projected element-element network with a probability
$\rho$ (bond percolation). When $\rho=0$ we recover the original mean-field
model where every module is completely dense. For large values of $\rho$, such as $\rho > 0.8$ or $0.9$, the modules
possess no intrinsic density above that of the overall density, and we recover a
non-modular random graph. What this means is that we now model modules with
approximately Erd\"os-R\'enyi graphs. In this pre-percolation phase, an element
fails if it loses all its neighbors. Such elements are removed from the
element-element network and they are counted as element failures towards the
modules.

\begin{figure}
    \centerline{\includegraphics[]{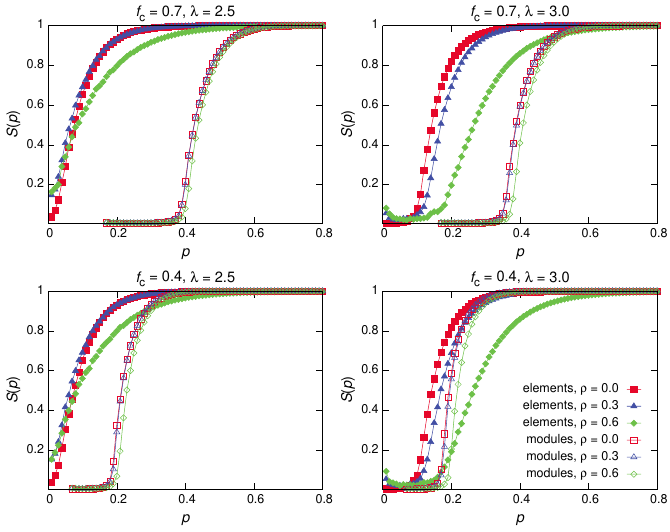}}
    \caption[Modules do not need to be completely dense]{
        Modules do not need to be completely dense. %
        Here we study the scale-free modular networks of Fig.~\ref{fig:scalefree}.
        However, before element failures occur we first delete links inside
        modules such that the final density of each module is approximately $1 -
        \rho$. When $\rho = 0$ we recover the results of Fig.~\ref{fig:scalefree}.  We
        consider two values of $\fc$ (top and bottom) and two scale-free
        exponents $\lambda$ (left and right). 
        Closed symbols correspond to element networks, open symbols to module
        networks.
        We see that in most cases the decreased density of modules has little
        effect. Only for the more extreme values of $\rho$ do we see a change, 
        which is reasonable because high values of $\rho$
        effectively destroy the modular nature of the network.
    \label{fig:preperc_delta3_powerlaw}}
\end{figure}

We study the robustness of these networks in
Fig.~\ref{fig:preperc_delta3_powerlaw}. We observe that $\rho$ has little if any
influence on the relative robustnesses of the two networks, over a range of
parameters. This provides further evidence that our results do not
pathologically depend on the mean-field nature (all-to-all coupling) of the underlying model.
We return to the importance of module density when we discuss our empirical results in Sec.~\ref{sec:empirical}.

\subsection{Additional models of module failures}
\label{subsec:additionalModelsOfModuleFailures}

We proposed a basic, mean-field model for how modules can fail. Instead of
considering the microscopic details of each module, we assumed that a module requires a
critical fraction $\fc$ of its original member elements to remain, regardless of
which particular elements actually remain. 
This was done for the sake of analytic tractability, yet there are numerous
other models one can study, typically of greater complexity. For example,
certain modules may possess distinct critical structure in such a way that the
module may continue to function so long as one key element remains, regardless
of how many other elements have failed. But failure of that one element will
invariably cause the module to fail.

\begin{figure}
    \centerline{
        {\includegraphics[width=0.5\textwidth,trim=3 0 1 0]{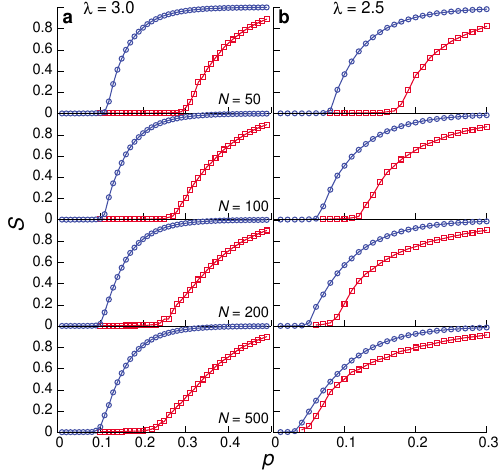}}
        \hspace*{-0.7em}
        {\includegraphics[width=0.5\textwidth,trim=3 0 1 0]{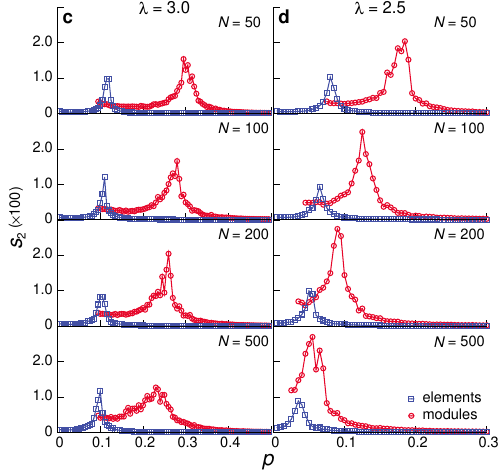}}
    }
    \caption[An alternative failure mode for modular networks]{ %
        An alternative failure mode for modular networks. %
        Here modules fail when they have fewer than $x=4$ original member
        elements remaining. This absolute cutoff size differs from the relative
        cutoff model where a fraction $\fc$ of the original
        member elements needed to remain for the module to remain functional.
        We see that, despite this very different mechanism for module failure,
        the difference in critical points between the two networks remains,
        supporting the generality of our results.
        \lett{a} %
        The size of the largest connected component for the element-element and
        module-module networks for a scale-free distribution of module sizes
        with scale-free exponent $\lambda = 3$. We see that the critical points
        remain different as we increase the maximum module size cutoff $N$.
        \lett{b} As per panel a but for a broader module size distribution
        ($\lambda = 2.5$). We see that increasing $N$ can decrease the gap
        between the element and module critical points.
        \lett{c-d} %
        The same networks as in panels a and b but now we plot the size of the
        \textit{second} largest connected component. This takes a maximum value
        at approximately the transition point $\pc$ and may more clearly illustrate
        how the critical points change as $N$ is varied.
        \label{fig:delta_powerlaw_absoluteFail}
    }
\end{figure}

As a first step towards exploring such alternative models of module robustness, consider the following.
Instead of requiring a critical fraction $\fc$ of elements for a
module to function, we now consider a module to remain if it has at least
$x>0$ elements remaining. This absolute number is now independent of the
original size of the module. We can study this analytically using nearly
the same calculation presented in Sec.~\ref{subsec:modulenetwork}; we need only replace $x(n) =
\lceil n\fc \rceil$ with $x=\mathrm{const}$. In
Fig.~\ref{fig:delta_powerlaw_absoluteFail} we study the robustness of the
scale-free modular networks considered in Fig.~\ref{fig:scalefree} under this new failure
criterion. We see that, while the critical point of the module-module network
does vary, it remains different from the critical point of the corresponding
element-element network for most parameter values. The primary change being that
increasing the maximum module size does improve the robustness of the module
network, when the scale-free exponent $\lambda<3$.

\begin{figure}
    \centerline{\includegraphics[]{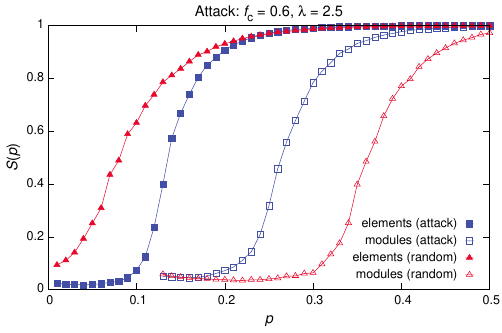}}
    \caption[Attacking modular networks]{Attacking modular networks. %
    Here we consider the robustness of scale-free modular networks (compare to
    Fig.~\ref{fig:scalefree}) under intentional attacks where element failures are no
    longer uniformly random, but instead the probability for an element to fail
    is proportional to the degree of the element in the original network such
    that higher degree elements are more likely to fail than lower degree
    elements.  We observe that, while smaller for attacks than random failures,
    the difference between the percolation critical points of the two networks
    remains. Simulations used $10^4$ elements.  
    \label{fig:attack_delta3_powerlaw}
    }
\end{figure}

In addition to random failure, seminal network studies also considered
\emph{attacks} where certain nodes---the high-degree hubs---are more likely to
fail. We again study the robustness of the scale-free network systems in Fig.~\ref{fig:scalefree}  but now elements fail
with probability proportional to their degree in the element-element network. We
see in Fig.~\ref{fig:attack_delta3_powerlaw} that these attacks do shift the
locations of both networks' percolation critical points, as expected. Yet, the
module-module network remains less robust to attacks than the element-element
network. Thus our qualitative results remain for multiple failure methods,
including targeted attacks.

\section{Empirical results}
\label{sec:empirical}

We study failures in the following four biological and technological real-world datasets:
A metabolic network, a protein-protein interaction network, a network of web pages captured by a web crawler, and a brain network
captured from fMRI data
The Metabolic and Protein-Protein Interaction (all) networks were previously used in Ahn \etal{} (2010); details are available there. The World Wide Web network is constructed from a web crawl made available by Google\footnote{Google no longer hosts these data, but it remains available at
\url{http://snap.stanford.edu/data/web-Google.html}.}. 
Finally, the Brain network was derived using normal subject fMRI data where each
node is a ``voxel'' dividing the brain spatially and links exist between voxels
with correlated time series. 
Full technical details can be found in M{\o}rup \etal{}~\shortcite{morup2010infinite}.
(We preprocessed the fMRI network to remove spurious connections; see
Appendix~\ref{app:brain} for full details.)

\begin{figure}
\centering
{\includegraphics[width=0.7\textwidth]{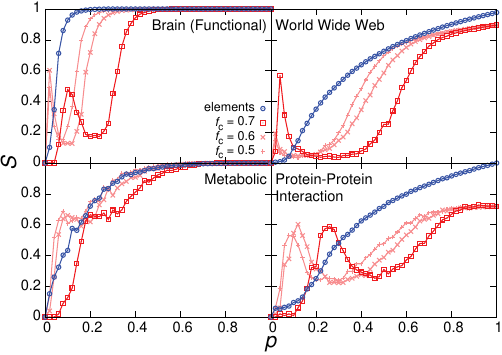}}\hfill
{\includegraphics[width=0.7\textwidth]{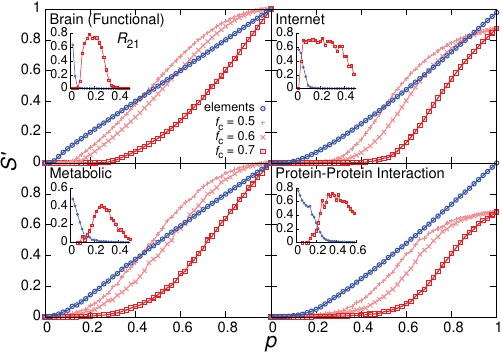}}
\caption{\coloronline{}
    Failures in a number of real, modular networks.  Many of these networks are
    robust to random failures (the element networks exhibit very small $\pc$).
    The behavior of the largest connect component for all the empirical networks
    qualitatively matches that of the model, as the identified modules uncouple
    faster than the network itself. (\textbf{top}) Our original definition of $S$ as the fraction
    of remaining nodes within the LCC tends to mask the transition for very small networks, as seen by the upward turn at small $p$ for some of the red curves. (\textbf{bottom}) We additionally plot $S^{\prime}$, the fraction of \emph{original} nodes in the LCC. This leads to a less dramatic transition but also avoids the denominator of $S$ becoming very small. (\textbf{insets}) Finally, to clearly demonstrate the robustness gap between the two networks, we also show the ratio $R_{21}$ of the size of
    the second largest to largest component ($\fc=0.7$) as a function of $p$, which
    tends to peak at $\pc$, further illustrating the difference in critical
    points for the two networks.  
    \label{fig:realworld}} 
\end{figure}

Unlike the analytical models (Sec.~\ref{sec:models}), here we do not know the modules in advance, so we estimate them
with an overlapping community detection algorithm~\cite{ahnbagrowlehman10}.
This algorithm works by extracting \emph{link communities} at the level of maximum partition density \cite{ahnbagrowlehman10}, which were then converted to overlapping node communities to provide the estimated modules. Only communities with at least three nodes were considered~\cite{ahnbagrowlehman10}.
These networks tend to be smaller than those previously discussed, introducing
finite-size effects that mask the behavior of $S$ (Fig.~\ref{fig:realworld}).  To overcome this, we additionally present
$S^{\prime}$, the fraction of \emph{original} nodes that remain in the largest connected
component. This slightly different definition behaves better for very small networks and high failure probabilities (small $p$) because the denominator does not go to zero, but the transition at the critical point is not as dramatic, making it harder to find the critical point.
To more clearly demonstrate that the critical points of element and module networks in the empirical data are not the same, we also calculate $R_{21}$, the ratio of the size
of the second largest to largest component ($R_{21}(p)$ tends to peak at the
critical point).
As shown in Fig.~\ref{fig:realworld}, the modules fall apart more easily than
the elements, qualitatively matching our model across a broad range of networks. 

\begin{figure}
    \centerline{\includegraphics[width=\textwidth]{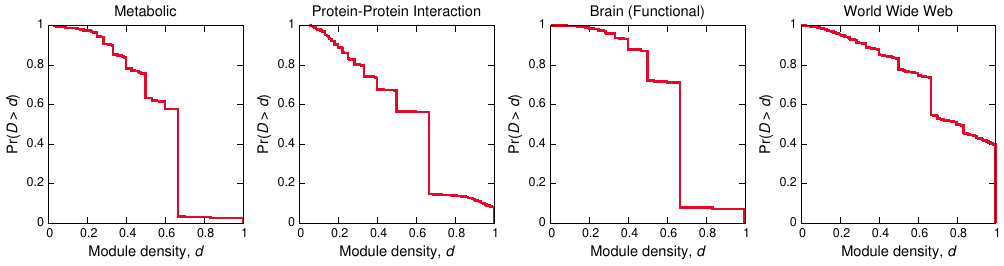}}
    \caption[Densities of empirically measured modules]{
        \textbf{Densities of empirically measured modules.}
        For each of the four empirical datasets, we compute the distribution of
        densities for subgraphs corresponding to the modules found using link
    communities~\cite{ahnbagrowlehman10}.
    The density $d$ is defined as the fraction of possible edges actually found
    within the module, i.e.\ $m / \binom{n}{2}$ where $m$ is the actual number
    of links in the module and $n$ is the number of nodes. 
    We see that few modules are completely dense, except for the world wide web
    network where approximately 40\% of modules are fully dense. Instead most
    network modules have densities between $d=0.2$ and $2/3$. This means that the
    empirical results shown in Fig.~\ref{fig:realworld}  do not require fully dense, mean-field
    modules for lower modular robustness to be present, further supporting the
    generality of our results.
    \label{fig:distrModuleDensities}
    }
\end{figure}

In Sec.~\ref{subsec:additionalModelsModules} and Fig.~\ref{fig:preperc_delta3_powerlaw} we discussed a model of modules that relaxed the need for them to be completely dense, and showed that the
robustness gap between the element network and module network remained.
That we do not require completely dense modules is further supported by the
evidence  presented here using real-world networks and their community structure.
These module networks are not created by projecting a bipartite node-community graph.
The modules themselves are not modeled but instead arise naturally from a community
detection method. These detection methods attempt to find dense graphs, but do
not impose structural restrictions. In Fig.~\ref{fig:distrModuleDensities} we
present the distribution of module densities---defined as the fraction of
potential links within a module that actually exist---for all modules of each
empirical network.  We see that most modules are dense, but not completely
dense: most modules have about two-thirds link density, corresponding to a value
of $\rho=1/3$ in our relaxed model.
Since the empirical networks show qualitatively similar relative robustness,
this further supports the fact that our results depend on the presence of dense
modules but not on strict forms for those modules.

\section{Discussion}
\label{sec:discuss}

We have used an analytically tractable network model to study the robustness of
modular networks to the random failures of elements. By analyzing a second network
detailing the connectivity between modules, we have shown that the overlapping
modular structure of the network is more susceptible to random failures than
expected. 
As mentioned previously, this modular network model makes mean-field assumptions
about both the nature of the modules and how element failures lead to module
failures. To understand whether these assumptions limit our results, we studied different models of modules within networks (Sec.~\ref{subsec:additionalModelsModules}) and
different types of module failure mechanisms (Sec.~\ref{subsec:additionalModelsOfModuleFailures}) including targeted
attacks (Fig.~\ref{fig:attack_delta3_powerlaw}). In all cases the presence of modules within the network
affects the robustness of the system.

There are a number of interesting avenues for further work.  We considered the
simplest case of random failures but further analysis of purposeful attacks
(where, e.g., elements with more connections are more likely to fail) are also
important.  Likewise, the model we use assumes that all links exist within
modules, but links between modules are certainly possible.  These additional
``weak'' links can only enhance the robustness of the element network, but will
not strengthen the module network, so that the network's functional resilience
does not improve.
Beyond structural characteristics of these modular networks it is important to
understand the effect of failures and modular structure on critical phenomena
such as synchronization~\cite{Arenas200893}, contact
processes~\cite{PhysRevLett.94.178701},
cascades~\cite{PhysRevLett.93.098701,buldyrev-havlin-catastrophic-cascade-failures-nature-2010,Brummitt20032012}
or other dynamics~\cite{RevModPhys.80.1275}.

Finally, this work may also help understand how analyses of empirical networks
are affected by missing data, of critical importance when finding communities,
or empirically discovering modules~\cite{Girvan11062002}. Here $p$ is the
probability that a network element is successfully captured by an experiment,
such as a high-throughput biological assay or web crawler, and the ``failure'' of
a module is now the inability of a hypothetical or idealized community detection
method to discover it due to the module's lack of density in the sampled
network. In this scenario, our results---the difference between the critical
points of the element and module networks---may indicate that, if the network is
sampled down to the intermediate regime where nodes are connected but modules
are uncoupled, the community overlap in the network will be underestimated,
allowing even non-overlapping community methods to succeed. Of course, this is
a simplified picture and requires further investigation. We do not know the true community structure and the networks are
likely to be already missing data. Yet, since the existence of strongly
overlapping community structure has been established in many networks (e.g.
Ahn \etal, 2010) and as we have shown that sampling tends to reduce overlap
between modules, we argue that community overlap in real networks is likely to
be underestimated.

\appendix{}

\section{Brain network preprocessing} \label{app:brain}

The Brain network was derived using normal subject fMRI data where each node is
a ``voxel'' dividing the brain spatially and links exist between voxels whose
respective BOLD time series are correlated.  We begin with the top 200k most
correlated links, measured using Mutual Information \cite{morup2010infinite}. 
A single voxel had very high degree, $k = 0.73N$ (the next
highest degree is $k = 0.096N$) so we first remove it. This leaves 5038 nodes
and 196311 links.  

\begin{figure}
  \centering
  {\includegraphics[]{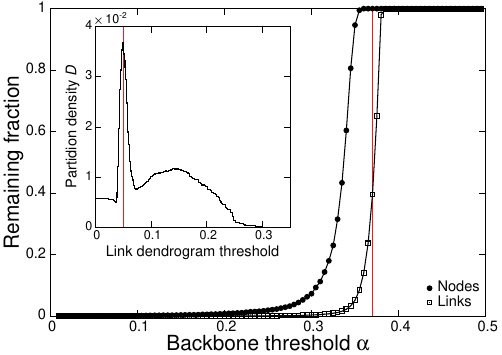}}
    \caption[The multiscale backbone and link communities of the
    fMRI brain network]{\small Extracting the multiscale backbone and link
        communities of the fMRI brain network.  We track the fraction of
        nodes and links remaining in the network as a function of the
        backbone threshold $\alpha$.  Choose $\alpha$ too small and little
        of the network remains; too big and the density is not altered.  We
        see a small window near $0.35 < \alpha < 0.4$ where the number of
        links drops but the majority of nodes remain.  We choose
        $\alpha=0.37$ (indicated) to exploit this. (\textbf{inset}) Partition density
        \cite{ahnbagrowlehman10} as a function of link dendrogram threshold
        for the extracted network.  The vertical line denotes the threshold 
        at which the dendrogram was cut to determine link communities.
    \label{fig:plot_backbone}
    } 
\end{figure}

We further preprocess this dense, weighted network by extracting its
\emph{multiscale backbone} \cite{serrano2009extracting}. To do so, we use the
Serrano algorithm~\shortcite{serrano2009extracting} with local heterogeneity
significance threshold $\alpha=0.37$. 
To determine this value of $\alpha$ we use the following approach.  The goal of
the backbone extraction method is to prune potentially spurious links by finding
significant links while disconnecting few nodes from the network. If $\alpha$ is
too small many nodes will lose all their neighbors since few links will be
significant. Yet if $\alpha$ is too high few links will be pruned since most
links will appear to be significant. Therefore we wish to choose $\alpha$ such that the
density of links is decreased but few nodes have been removed. 
In Fig.~\ref{fig:plot_backbone} we plot the fraction of nodes and the fraction
of links remaining in the graph as a function of $\alpha$. Indeed we see a
distinct window $0.35 < \alpha < 0.39$ where link removals occur but few nodes
have been lost. We choose $\alpha=0.37$, a value in the middle of this range
where many links have been removed but nearly all nodes are still present in the
network.

After extracting significant links using the backbone algorithm, the fMRI data
is reduced to a final network of 5038 nodes and 77680 links.  
For the brain network (and all networks), link communities were extracted at the link dendrogram
level of maximum partition density~\cite{ahnbagrowlehman10}, providing the
estimated modules.  As in Ahn, \etal{}~\shortcite{ahnbagrowlehman10}, only communities with at
least three nodes were considered. In Fig.~\ref{fig:plot_backbone} (inset) we plot the partition
density of the Brain network as a function of the height (or threshold) of the
link dendrogram (see Ahn, \etal{} for details~\shortcite{ahnbagrowlehman10}). We see
a sharp peak at a threshold near the root of the tree, giving a clear indicator
for the most modular component of the network's link hierarchy.

\section*{Acknowledgments}

We thank H.~Rozenfeld, F.~Simini, Y.-R.~Lin, J.~Menche, D.~Wang,
D.~ben-Avraham, and A.-L.~Barab\'asi for many useful discussions; and
H.~Siebner and K.~Madsen at Hvidovre Hospital's Danish Research Centre for
Magnetic Resonance for normal-patient fMRI data.  The authors acknowledge the
Center for Complex Network Research, supported by the James S.~McDonnell
Foundation, the NSF, NIH, US ONR and ARL, DTRA, and NKTH NAP.


\end{document}